\documentclass[pra,twocolumn,showpacs,preprintnumbers,amsmath,amssymb]{revtex4}
\usepackage{mathrsfs}
\usepackage{bbm}
\usepackage{amsfonts}
\usepackage{tipa}

\usepackage{epsfig,graphicx}
\usepackage{amstext}
\usepackage{amsmath}
\usepackage{graphicx}
\usepackage{times}

\begin{document}


\title{Upper bound and shareability of quantum discord based on
       entropic uncertainty relations}

\author{Ming-Liang Hu$^{1}$}
\email{mingliang0301@163.com}
\author{Heng Fan$^{2}$}
\email{hfan@iphy.ac.cn}
\affiliation{$^{1}$School of Science, Xi'an University of Posts and
               Telecommunications, Xi'an 710121, China \\
             $^{2}$Beijing National Laboratory for Condensed Matter Physics,
               Institute of Physics, Chinese Academy of Sciences, Beijing
               100190, China}

\begin{abstract}
By using the quantum-memory-assisted entropic uncertainty relation
(EUR), we derive a computable tight upper bound for quantum discord,
which applies to an arbitrary bipartite state. Detailed examples
show that this upper bound is tighter than other known bounds in a
wide regime. Furthermore, we show that for any tripartite pure
state, the quantum-memory-assisted EUR imposes a constraint on the
shareability of quantum correlations among the constituent parties.
This conclusion amends the well accepted result that quantum discord
is not monogamous.
\end{abstract}


\maketitle

Quantum correlations are of special importance in quantum
information processing, such as in the deterministic quantum
computation with one qubit \cite{Datta}, and other related quantum
protocols \cite{Dakicnp,gumile,Modirmp}. Among different measures of
quantum correlation, quantum discord (QD) \cite{Ollivier} has been
attracting particular attention. Various aspects of QD , e.g., the
role it played in identifying quantum phase transition \cite{Qpt},
its local creativity \cite{creation}, and operational interpretation
\cite{Madhok}, have been explored. Its peculiar behaviors in
evolution under noisy environments \cite{sudden1,sudden2} have also
been investigated.

Despite the significance, the value of QD is notoriously difficult
to calculate due to the optimization procedure involved. Analytical
results are known only for certain special classes of states
\cite{Luoetal,Cenlx,Chitambar}. Particularly, it has been proved
that it is impossible to obtain a closed expression for QD, even for
general states of two qubits \cite{Girolami}. This fact makes it
desirable to obtain some computable bounds for QD, and several
attempts have been devoted to this issue in the past few years
\cite{bound1,Xizj,Yusx,Zhangc}.

In this work, we reexamined the above issue from some alternative
perspectives. We noted that the quantum correlation plays a
deterministic role in improving the prediction precision of an
imaginary ``uncertainty game'' \cite{Berta}, which has been further
explored in several recently published papers \cite{uncer}.
Particularly, a connection between entanglement and measurement
uncertainty was established in a very recent work \cite{Berta2}.
Here, instead of concentrating on the role that QD played in
tightening the lower bound of the new entropic uncertainty relation
(EUR) \cite{Pati}, we reversely consider how this EUR constrains the
magnitude of QD. Our study shows that from the uncertainty principle
represented as the EUR, one can derive certain improved upper bounds
for QD. These bounds are tighter in a wide regime than those
obtained in the literature \cite{bound1,Xizj}.

Another issue we will study is how the EUR affects shareability of
quantum correlations among different subsystems. It is well known
that QD does not satisfy the monogamy relation which is considered a
fundamental property concerning the resource shareability among
multi-parties \cite{monogamy}. Naturally, a question arises as to
whether there exists any constraint on the shareability of QD. We
find that the EUR sets a fundamental limit on the shareability of QD
for all the tripartite pure states. This can be considered as an
amendment to the fact that QD violates the monogamy condition.

Let us first recall the definition of QD, which is based on the
partition of the total correlations in a state $\rho_{AB}$, measured
by the quantum mutual information
$I(\rho_{AB})=S(\rho_A)+S(\rho_B)-S(\rho_{AB})$, into two different
parts, i.e., the classical part and the quantum part. The classical
part $J_A(\rho_{AB})$, also known as the classical correlation
\cite{Ollivier}, is defined as
\begin{equation}\label{eq1}
 J_A(\rho_{AB})=S(\rho_B)-\min_{\{E_k^A\}}S(B|\{E_k^A\}),
\end{equation}
where $S(\rho_B)=-{\rm Tr}(\rho_B\log_2\rho_B)$ denotes the von
Neumann entropy of the reduced density operator $\rho_{B}={\rm
Tr}_{A}\rho_{AB}$, and $S(B|\{E_k^A\})=\sum_{k}p_{k}S(\rho_{B|k})$
is the averaged conditional von Neumann entropy of the nonselective
postmeasurement state $\rho_{B|k}={\rm Tr}_{A}(E_k^A\rho_{AB})/p_k$
after the positive operator valued measure (POVM) on party $A$, with
$p_k=\text{Tr}(E_k^A\rho_{AB})$.

The quantum part $D_A(\rho_{AB})$, which is QD under our
consideration, is then obtained by subtracting $J_A(\rho_{AB})$ from
$I(\rho_{AB})$ \cite{Ollivier}, namely,
\begin{equation}\label{eq2}
 D_A(\rho_{AB})=\min_{\{E_k^A\}}S(B|\{E_k^A\})-S(B|A),
\end{equation}
where $S(B|A)=S(\rho_{AB})-S(\rho_A)$ denotes the conditional von
Neumann entropy of $\rho_{AB}$. For general $\rho_{AB}$, a tight
upper bound for QD is proven to be \cite{bound1}
\begin{equation}\label{eq3}
 D_A(\rho_{AB})\le S(\rho_A),
\end{equation}
with equality holding if and only if the complex Hilbert space of subsystem
$B$ can be decomposed as $\mathcal {H}_B=\mathcal {H}_{B^L}\otimes
\mathcal {H}_{B^R}$ such that
$\rho_{AB}=|\psi\rangle_{AB^L}\langle\psi|\otimes\rho_{B^R}$
\cite{Xizj}.

The quantum-memory-assisted EUR was initially conjectured by Renes
and Boileau \cite{Renes} and then proven by Berta {\it et al.}
\cite{Berta}. It reads
\begin{equation}\label{eq4}
 S(Q|B)+S(R|B)\ge \log_2 \frac{1}{c}+S(A|B),
\end{equation}
where $S(X|B)$ is the conditional entropy of the postmeasurement
state $\rho_{XB}=\sum_{k}(|\psi_k^X\rangle\langle\psi_k^X|\otimes
\mathbb{I})\rho_{AB}(|\psi_k^X\rangle\langle\psi_k^X|\otimes
\mathbb{I})$, with $|\psi_k^X\rangle$ being the eigenvectors of
$X=\{Q,R\}$. Moreover, $c$ in Eq. \eqref{eq4} quantifies the
incompatibility of the observables $Q$ and $R$. It is defined as
$c=\max_{k,l}|\langle \psi_k^Q|\psi_l^R\rangle|^2$ . Here, the
subsystem $B$ is called a quantum memory, as it stores information
which can be used by one player of the uncertainty game to infer the
measurement outcome of his counterpart \cite{Berta}.

Experimentally, the EUR of Eq. \eqref{eq4} has been tested in
systems of photon pairs \cite{Licf}, and is proposed for testing in
nitrogen-vacancy (NV) center in diamond \cite{apl}. Theoretically, a
tighter lower bound of measurement uncertainty than that presented
in the right-hand side (RHS) of Eq. \eqref{eq4} is obtained in a
recent work \cite{Pati}. By incorporating the discrepancy between QD
and the classical correlation into account, the following inequality
is proven:
\begin{equation}\label{eq5}
 S(Q|B)+S(R|B)\ge \log_2 \frac{1}{c}+S(A|B)+\max\{0,-\Delta\},
\end{equation}
where $\Delta=J_A(\rho_{AB})-D_A(\rho_{AB})$ characterizes the
imbalance between the classical correlation and QD \cite{Felipe}.
Therefore, the lower bound of Berta {\it et al.} [Eq. \eqref{eq4}]
is tightened whenever $\Delta<0$, i.e., when the quantum correlation
in the joint system of the quantum memory and the measured particle
exceeds the classical correlation that exists in the same system.

We remark here that the correlation discrepancy $\Delta$ equals
$I(\rho_{BC})-2E_f(\rho_{BC})$ when one takes the purified state
$|\Psi\rangle_{ABC}$ for $\rho_{AB}$ into consideration
\cite{Felipe}, where $E_f(\rho_{BC})=\min\sum_i p_i S({\rm Tr}_C
|\psi_i\rangle_{BC}\langle\psi_i| )$ represents the entanglement of
formation (EoF) \cite{Bennett} for state $\rho_{BC}={\rm Tr}_A(
|\Psi\rangle_{ABC}\langle\Psi|)$, and the minimum is taken over all
the pure-state decompositions $\rho_{BC}=\sum_i p_i
|\psi_i\rangle_{BC}\langle\psi_i|$. Meanwhile,
$I(\rho_{BC})-2E_f(\rho_{BC})$ is also found to be equal to
$\tau_D=D_A(\rho_{A:BC})-D_A(\rho_{AB})-D_A(\rho_{AC})$ \cite{Ren}.
$\tau_D$ is the discord monogamy score introduced in \cite{score}.
Therefore, we have $\Delta=\tau_D$, which indicates that the lower
bound of the EUR in Eq. \eqref{eq4} is improved whenever the
purification $|\Psi\rangle_{ABC}$ for $\rho_{AB}$ violates the
monogamy inequality $D_A(\rho_{AB})+D_A(\rho_{AC})\le
D_A(\rho_{A:BC})$, i.e., $\Delta<0$ whenever $|\Psi\rangle_{ABC}$ is
not monogamous.

With the above preliminaries, we now show applications of the EUR
\eqref{eq5} in deriving improved upper bounds on QD. To this end,
and for the purpose of showing figures of merit using this method,
we first introduce a slightly stronger upper bound of QD than that
presented in Eq. \eqref{eq3}, which is given by
\begin{eqnarray}\label{eq8}
 D_A(\rho_{AB})\le \min\{S(\rho_A),S(\rho_A)-S(A|B)\}.
\end{eqnarray}

\begin{figure}
\centering
\resizebox{0.35\textwidth}{!}{%
\includegraphics{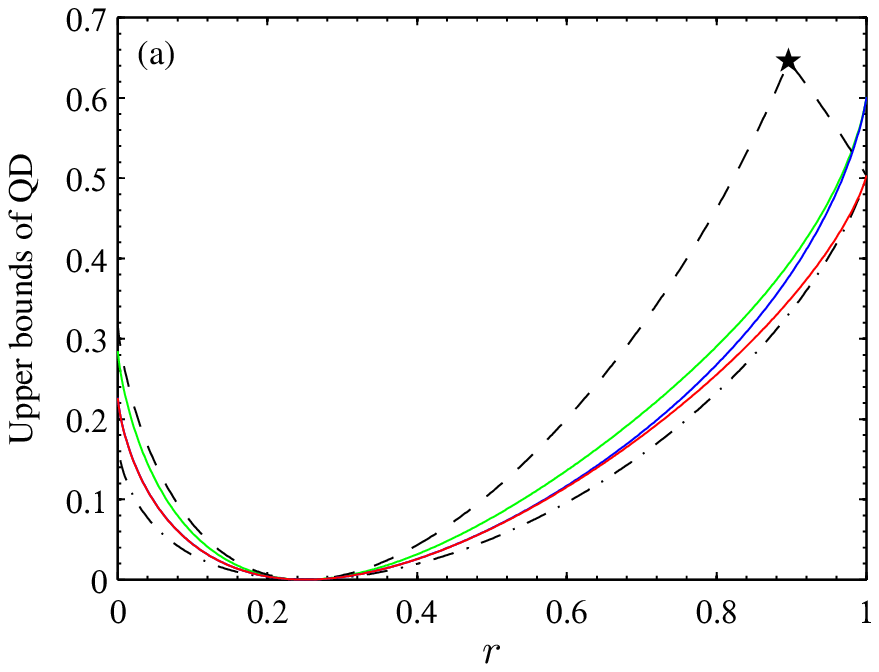}}
\centering
\resizebox{0.35\textwidth}{!}{%
\includegraphics{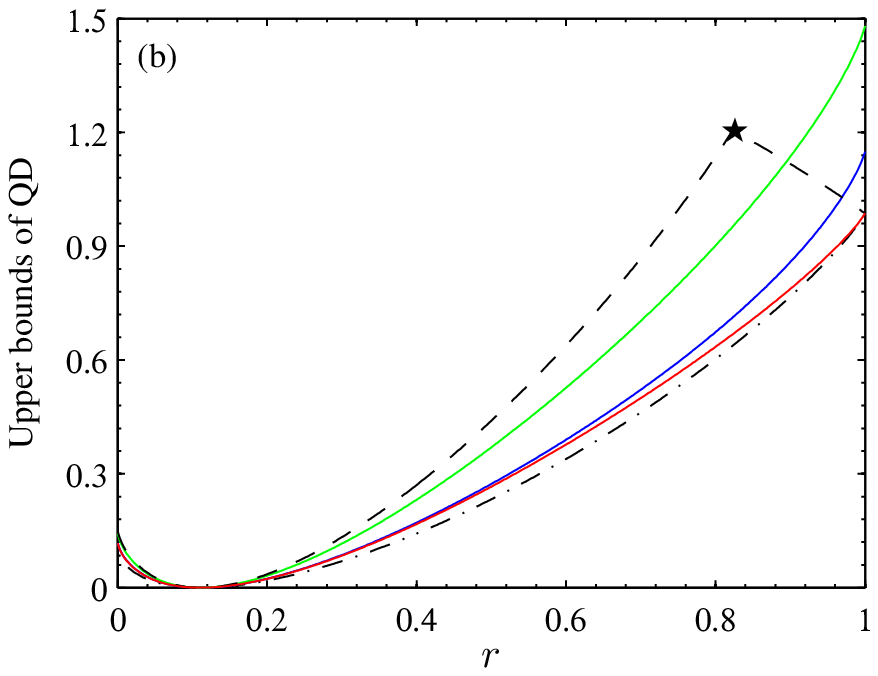}}
\caption{(Color online) Upper bounds of QD for $\rho_{\rm PP}$ of
Eq. \eqref{eq9} with (a) $d=2$, $u_1=2\sqrt{2}/3$, $u_2=1/3$ and (b)
$d=3$, $u_1=\sqrt{7}/3$, $u_{2,3}=1/3$. The dash-dotted and the
dashed lines are the exact results of QD and its upper bounds given
by Eq. \eqref{eq8} [the stars denote critical points after which
$S(\rho_A)$ in  Eq. \eqref{eq8} dominates], while the solid red,
blue, and green lines (from bottom to top) are those given by
$\Lambda_\alpha$ [Eqs. \eqref{eq10} and \eqref{eq12}] with
$\alpha={\rm T}$, ${\rm M}$, and ${\rm F}$, respectively. }
\label{fig:1}
\end{figure}

This upper bound tightens that presented in Eq. \eqref{eq3} for all
$\rho_{AB}$ with positive conditional entropy $S(A|B)$
[$D_A(\rho_{AB})\le S(\rho_A)-S(A|B)=I(\rho_{AB})$ holds obviously
true by its definition]. This occurs for several natural bipartite
states; see, for example, the dashed lines in Fig. \ref{fig:1}
obtained by Eq. \eqref{eq8} for the $d\otimes d$ pseudopure state
\cite{Chitambar}
\begin{eqnarray}\label{eq9}
 \rho_{\rm PP}=\frac{1-r}{d^2-1}\mathbb{I}+\frac{rd^2
 -1}{d^2-1}|\psi\rangle\langle\psi|,
\end{eqnarray}
where $|\psi\rangle=\sum_{i=1}^{d}u_i|ii\rangle$, with
$\sum_{i=1}^{d}u_i^2=1$. For both plots displayed in Fig.
\ref{fig:1}, the dashed lines before the sudden change points
$r_{SC}$ denoted by the stars correspond to the upper bounds of
$D_A(\rho_{AB})$ given by $S(\rho_A)-S(A|B)$, while after the points
$r_{SC}$, they are given by the original bound $S(\rho_A)$. Due to
the high symmetry of $\rho_{\rm PP}$, analytical results of the QD
can also be obtained \cite{Chitambar} and they are denoted by the
dash-dotted lines illustrated in Fig. \ref{fig:1}.

Based on the result of Eq. \eqref{eq8}, we now present our improved
upper bound to QD via the following theorem.

\emph{Theorem 1.} For any bipartite state $\rho_{AB}$, the QD
satisfies
\begin{eqnarray}\label{eq10}
 D_A(\rho_{AB}) \le \min \{S(\rho_A),I(\rho_{AB}), \Lambda_{\rm T}\},
\end{eqnarray}
where $\Lambda_{\rm T}=[\delta_{\rm T}+I(\rho_{AB})]/2$, and
\begin{eqnarray}\label{eq11}
 \delta_{\rm T}=S(Q|B)+S(R|B)-\log_2 \frac{1}{c}-S(A|B),
\end{eqnarray}
which characterizes the discrepancy between uncertainty of the
measurement outcomes of $Q$ and $R$ (inferred from projective
measurement on $A$ and quantum state tomography on $B$, known as the
tomographic estimate in \cite{Licf}) and its lower bound depicted on
the RHS of Eq. \eqref{eq4}.

\emph{Proof.} Due to Eq. \eqref{eq8}, it suffices to prove that (i)
the inequality $D_A(\rho_{AB})\le \Lambda_{\rm T}$, and (ii) it is
possible for $\Lambda_{\rm T}$ to be smaller than or equal to
$S(\rho_A)$ and $I(\rho_{AB})$.

The first one, that is, $D_A(\rho_{AB})\le \Lambda_{\rm T}$, can be
shown to be true by reexpressing $\Delta$ in Eq. \eqref{eq5} as
$I(\rho_{AB})-2D_A(\rho_{AB})$, which gives immediately $\delta_{\rm
T} \ge 2D_A(\rho_{AB})-I(\rho_{AB})$, and therefore
$D_A(\rho_{AB})\le \Lambda_{\rm T}$.

The second one can be proven by taking the minimum uncertainty
states of Berta {\it et al.} [e.g., the Greenberger-Horne-Zeilinger
state $(|000\rangle+|111\rangle)/\sqrt{2}$; refer to \cite{Coles}
for more details about this kind of state] as an example, which
correspond to $\delta_{\rm T}=0$. Therefore, the requirement (ii)
reduces to $I(\rho_{AB})/2\le S(\rho_A)$ and $I(\rho_{AB})/2\le
I(\rho_{AB})$. The former one can always be satisfied due to the
Araki-Lieb inequality $|S(\rho_A)-S(\rho_B)|\le S(\rho_{AB})$
\cite{Nielsen} and the latter one is obvious. This completes the
proof.\hfill{$\blacksquare $}

One can make the upper bound $\Lambda_{\rm T}$ better by choosing
appropriate observables. Particularly, when $Q$ and $R$ are
complementary such that $\log_2(1/c)=\log_{2}d_A$, with $d_A$ being
the dimension of $\mathcal {H}_A$, the upper bound $\Lambda_{\rm T}$
of $D_A(\rho_{AB})$ in Eq. \eqref{eq10} is saturated for the
isotropic state of arbitrary dimensions \cite{isotropic}, i.e., for
all $\rho_{\rm PP}$ of Eq. \eqref{eq9} with $u_i=1/\sqrt{d}$. Thus
the upper bound $\Lambda_{\rm T}$ we obtained is tight.

For general $\rho_{\rm PP}$, from the solid red lines shown in Fig.
\ref{fig:1} which are obtained by choosing the observables $Q$ and
$R$ such that $\log_2(1/c)=\log_2 d$ [note that for $\rho_{\rm PP}$
in Eq. \eqref{eq9}, $d_A=d_B=d$], one can see obviously that the
upper bound $\Lambda_{\rm T}$ tightens that given by Eq.
\eqref{eq8}, and for the special cases of $r=1/d^2$ [with
$D_A(\rho_{AB})=0$] and $r=1$ [with
$D_A(\rho_{AB})=-\sum_{i=1}^{d}u_i^2 \log_2 u_i^2$], the upper bound
$\Lambda_{\rm T}$ is saturated.

Moreover, by using the facts that projective measurements increase
entropy \cite{Nielsen} and Fano's inequality \cite{Berta,Licf}, one
can obtain two slightly weaker bounds for QD as follows:
\begin{eqnarray}\label{eq12}
 D_A(\rho_{AB}) \le \min \{S(\rho_A),I(\rho_{AB}),\Lambda_\alpha\},
\end{eqnarray}
where $\Lambda_{\alpha}=[\delta_{\rm \alpha}+I(\rho_{AB})]/2$, with
$\alpha={\rm \{M,F\}}$. Similarly, $\delta_{\rm M}$ and $\delta_{\rm
F}$ can be obtained directly by replacing the first two terms on the
RHS of Eq. \eqref{eq11} with $S(Q|Q)+S(R|R)$ and
$h(p_Q)+h(p_R)+(p_Q+p_R) \log_2(d_A-1)$, respectively. Here,
$S(X|X)$ ($X=Q,R$) denotes the conditional von Neumann entropy of
the postmeasurement state $\rho_{XX}$ obtained via two-side
projective measurements on $\rho_{AB}$, and $h(p_X)$ is the binary
entropy of the probability distribution $p_X$ corresponding to
different outcomes of $X$ on $A$ and $X$ on $B$.

The upper bounds given in Eq. \eqref{eq12} are, in general, weaker
than that of Eq. \eqref{eq10} in that $\Lambda_{\rm M,F} \ge
\Lambda_{\rm T}$, but may be favored for their ease of experimental
accessibility \cite{Licf}. Particularly, they may still be tighter
than that given by Eq. \eqref{eq8} under certain circumstances. See,
for example, the solid blue (given by $\Lambda_{\rm M}$) and solid
green (given by $\Lambda_{\rm F}$) lines displayed in Fig.
\ref{fig:1} for the pseudopure states of Eq. \eqref{eq9}, which are
nearly overlapped during the small $r$ regions. Clearly, both of the
bounds described by $\Lambda_{\rm M}$ and $\Lambda_{\rm F}$ are
tighter than that given by Eq. \eqref{eq8} in most intervals of the
mixing parameter $r$.

At this stage, one may wonder what the other implications would be
of the tightened EUR in Eq. \eqref{eq5}. Here, we show that it also
implies a constraint on the shareability of QD among different
parties of a composite system.

\emph{Theorem 2.} For any tripartite state $\rho_{ABC}$ with
$S(\rho_A)=-S(A|BC)$, we have
\begin{eqnarray}\label{eq13}
D_A(\rho_{AB})+D_A(\rho_{AC}) \le D_A(\rho_{A:BC})+\delta_{\rm
T}.
\end{eqnarray}

\emph{Proof.} First, Eq. \eqref{eq5} means $\delta_{\rm T} \ge
D_A(\rho_{AB})-J_A(\rho_{AB})$. This, together with the inequality
$D_A(\rho_{AC})+J_A(\rho_{AB})\leq S(\rho_A)$ which is applicable
for arbitrary $\rho_{ABC}$ \cite{Koashi}, results in
\begin{eqnarray}\label{eq14}
D_A(\rho_{AB})+D_A(\rho_{AC}) \le  S(\rho_A)+\delta_{\rm T}.
\end{eqnarray}
Thus, Eq. \eqref{eq13} holds obviously for all the tripartite pure
states $|\Psi\rangle_{ABC}$ because we always have
$D_A(\rho_{A:BC})=S(\rho_A)$, and $S(\rho_A)=-S(A|BC)$. Moreover, we
know from Ref. \cite{Xizj} that even for mixed $\rho_{ABC}$,
$S(\rho_A)=-S(A|BC)$ if and only if there exists a factorization
$\mathcal {H}_{BC}=\mathcal {H}_{(BC)^L}\otimes \mathcal
{H}_{(BC)^R}$ for the Hilbert space $\mathcal {H}_{BC}$ such that
$\rho_{ABC}=|\psi\rangle_{A(BC)^L}\langle\psi|\otimes\rho_{(BC)^R}$,
and therefore
$D_A(\rho_{A:BC})=D_A(|\psi\rangle_{A(BC)^L})=S(\rho_A)$.
\hfill{$\blacksquare $}

The inequality \eqref{eq13} is a released version of the monogamy
relation of QD \cite{monogamy}. It applies for all tripartite pure
states and to extended classes of mixed states. As $\delta_{\rm T}$
is non-negative due to Eq. \eqref{eq4}, the inequality \eqref{eq13}
implies immediately that even if QD may violate the monogamy
inequality, the different subsystems of $ABC$ still cannot be freely
correlated. That is, subsystem $A$ cannot share an unlimited amount
of quantum correlations individually with both $B$ and $C$, as the
summation is limited by $ D_A(\rho_{A:BC})+\delta_{\rm T}$.
Therefore, we see that although the monogamy inequality of QD may be
violated, there exists a limitation for QD shareability.

Moreover, as a corollary, we emphasize here that Eq. \eqref{eq13}
also yields a sufficient condition for the monogamy of QD in a class
of tripartite states $\rho_{ABC}$ with $S(\rho_A)=-S(A|BC)$. That
is, \emph{there exist measurement operators $Q$ and $R$ such that
the discrepancy $\delta_{\rm T}$ defined in Eq. \eqref{eq11}
vanishes, i.e., $\delta_{\rm T}=0$}. This occurs, for instance, for
the reduced $\rho_{AB}$ being the minimum uncertainty state of Berta
{\it et al.} \cite{Coles}, namely, the bipartite states
$\rho_{AB}={\rm Tr}_C(|\Psi\rangle_{ABC}\langle\Psi|)$ saturate the
lower bound of Eq. \eqref{eq4}.

Finally, it is worthwhile to note that the original definition of QD
introduced by Ollivier and Zurek \cite{Ollivier} is measurement
dependent, thus there are two possible lines for studying the
monogamous character of QD. The case we discussed in Theorem 2
corresponds to that with the measurements being performed on the
same subsystem $A$. When we go forward along another line for which
the measurements were performed on different subsystems (some
discussions along this line can be found in
\cite{monogamy,Felipepra}), we can also establish a connection
between the EUR and shareability of QD.

\emph{Theorem 3.} For any tripartite pure state
$|\Psi\rangle_{ABC}$, we have
\begin{eqnarray}\label{eq15}
D_B(\rho_{AB})+D_C(\rho_{AC})
                        \le D_{BC}(\rho_{A:BC})+\bar{\delta}_{\rm T},
\end{eqnarray}
where $\bar{\delta}_{\rm T}=[\delta_{\rm T}^{(BA)}+\delta_{\rm
T}^{(CA)}]/2$, with
\begin{eqnarray}\label{eq16}
 \delta_{\rm T}^{(BA)}=S(Q_B|A)+S(R_B|A)-\log_2 \frac{1}{c}-S(B|A),\nonumber\\
 \delta_{\rm T}^{(CA)}=S(Q_C|A)+S(R_C|A)-\log_2 \frac{1}{c}-S(C|A).
\end{eqnarray}

\emph{Proof.} By making the substitutions $A\rightarrow X$ ($X=B$ or
$C$) and $B\rightarrow A$ to the EUR in Eq. \eqref{eq5}, we obtain
\begin{eqnarray}\label{eq17}
 D_X(\rho_{AX})\leq
               \frac{1}{2}[\delta_{\rm T}^{(XA)}+S(\rho_X)-S(X|A)],
\end{eqnarray}
for arbitrary $\rho_{ABC}$. Consequently,
\begin{eqnarray}\label{eq18}
D_B(\rho_{AB})+D_C(\rho_{AC}) &\le& S(\rho_A)+\bar{\delta}_{\rm T}\nonumber\\
                               &&   -\frac{1}{2}[S(A|B)+S(A|C)]\nonumber\\
                              &\le& S(\rho_A)+\bar{\delta}_{\rm T},
\end{eqnarray}
where the second inequality is due to the strong subadditivity of
the von Neumann entropy \cite{Nielsen}. Then, Eq. \eqref{eq15} is
obviously true for any $|\Psi\rangle_{ABC}$ because for pure states
we always have $D_{BC}(\rho_{A:BC})=D_A(\rho_{A:BC})=S(\rho_A)$.
\hfill{$\blacksquare $}

This theorem also implies that if there exist observables $Q$ and
$R$ giving nullity of $\bar{\delta}_{\rm T}$, then the state
$|\Psi\rangle_{ABC}$ will obey the monogamy of discord condition
$D_B(\rho_{AB})+D_C(\rho_{AC}) \le D_{BC}(\rho_{A:BC})$. It shows
again the power of the EUR for exploring monogamy properties of
discord.

In summary, we have shown applications of a generalized EUR
\cite{Pati} in obtaining an improved tight upper bound for QD. This
bound applies to bipartite states of arbitrary dimensions and
tightens that given in the literature \cite{bound1,Xizj}. In
addition, we have also shown applications of the EUR in identifying
an inequality which constrains the shareability of QD between
different parties of a composite system. More specifically, we
showed that even if QD may not respect the monogamy relation, the
quantum correlations still cannot be freely shared. As an amendment
to the violation of monogamy relation for QD, a released
monogamylike relation is still satisfied for all tripartite pure
states. We hope that these results may provide useful insights as to
what role quantum correlations play in the fundamental theory of the
uncertainty principle, as well as to how the uncertainty principle,
particularly those of the entropic forms, imposes constraints on the
strength and distributions of quantum correlations.
\\

This work was supported by NSFC (11205121, 10974247, 11175248), the
``973'' program (2010CB922904), NSF of Shaanxi Province
(2010JM1011), and the Scientific Research Program of the Education
Department of Shaanxi Provincial Government (12JK0986).

\newcommand{\NP}{Nat. Phys. }
\newcommand{\PRL}{Phys. Rev. Lett. }
\newcommand{\PRA}{Phys. Rev. A }
\newcommand{\PRB}{Phys. Rev. B }
\newcommand{\RMP}{Rev. Mod. Phys. }
\newcommand{\JPA}{J. Phys. A }
\newcommand{\JPB}{J. Phys. B }
\newcommand{\PLA}{Phys. Lett. A }
%

%

\end{document}